\documentstyle[prb,aps,epsf]{revtex} 

\def\ai{{\it ab initio}}	
\def\Ai{{\it Ab initio}}	
\def\FFH{\widehat{\rm FFH''}}	
\def\HFF{\widehat{\rm H'FF}}	
\def\R{{\bbox R\/}}		
\def\frac#1#2{{#1 \over #2}}	

\begin{document}
\draft
\title{Towards an effective potential for the monomer, dimer, hexamer, solid and
liquid forms of hydrogen fluoride}
\author{Raffaele Guido Della Valle}
\address{Dipartimento di Chimica Fisica e Inorganica, Universit\`a di Bologna, \\
Viale Risorgimento 4, I-40136 Bologna, Italy}
\author{Domenico Gazzillo}
\address{INFM Unit\`a di Venezia e Dipartimento di Chimica Fisica,\\
Universit\`a di Venezia, S. Marta 2137, I-30123 Venezia, Italy}
\date{\today}
\maketitle

\begin{abstract}
We present an attempt to build up a new two-body effective potential
for hydrogen fluoride, fitted to theoretical and experimental data
relevant not only to the gas and liquid phases, but also to the
crystal. The model is simple enough to be used in Molecular Dynamics
and Monte Carlo simulations. The potential consists of: a) an {\it
intra}-molecular contribution, allowing for variations of the
molecular length, plus b) an {\it inter}-molecular part, with three
charged sites on each monomer and a Buckingham ``$\exp\!-6$''
interaction between fluorines. The model is able to reproduce a
significant number of observables on the monomer, dimer, hexamer,
solid and liquid forms of HF. The shortcomings of the model are
pointed out and possible improvements are finally discussed.
\end{abstract}




\section{Introduction}

Hydrogen fluoride (HF) is one of the simplest molecules capable of
forming hydrogen bonds. Despite the simplicity of the compound, the
theoretical description of its behavior, especially in the liquid
phase, is still far from being fully satisfactory. Much relevant
theoretical work has concentrated on the problem of determining a
potential model suitable for computer simulations. For this purpose a
model is needed that reproduces correctly the main features of the
real interaction potential and which is simple enough to be computed
efficiently.

In the last two decades several empirical potentials have been developed for
Molecular Dynamics (MD) or Monte Carlo (MC) simulations of {\it liquid} HF.
\cite{Klein1979,Cournoyer1984,Jedlovszky1997a,Jedlovszky1997b,Sun1992,Jorgensen1978,Jorgensen1979,Honda1992}
Of these models, only the three-site ones
\cite{Klein1979,Cournoyer1984,Jedlovszky1997a,Jedlovszky1997b} are able to
reproduce correctly the dipole and quadrupole moments of the monomer. These
models represent the charge distribution of each monomer by fractional charges
placed at three sites on the molecular axis, two on the F and H nuclei and the
third one at an appropriate position X along the F-H bond. A first three-site
model (called HF3) was developed by Klein and McDonald \cite{Klein1979} by
fitting {\ai} results for the potential energy surface of the (HF)$_2$
dimer. Cournoyer and Jorgensen \cite{Cournoyer1984} proposed a second
three-site model (called HFC), with a simplified non-Coulombic part consisting
of a single Lennard-Jones interaction between the fluorines. The parameters of
the model were fitted directly to experimental thermodynamic data for the
liquid, while simultaneously providing a reasonable equilibrium geometry of
the dimer. Recently, Jedlovszky and Vallauri
\cite{Jedlovszky1997a,Jedlovszky1997b} presented two further models, hereafter
referred to as HF-JV1 and HF-JV2, respectively. HF-JV1 is a variant of HFC,
with charges reproducing both the monomer dipole and quadrupole and with an
accurate treatment of the long range part of the Coulombic interactions,
neglected in the original work on HFC. \cite{Cournoyer1984} The HF-JV2 model
includes molecular polarizability, by adding an induced point dipole moment at
each F site, while keeping the charge distribution of HF-JV1 unchanged. The
scalar polarizability of the molecules was set equal to its experimental
value, while the two parameters of the Lennard-Jones interaction between F
atoms were fitted to the experimental values of the liquid density and
internal energy.

Unfortunately, none of the available models is able to reproduce, in a fully
satisfactory way, both the thermodynamics and the structure of liquid and
gaseous HF. In the search for a new potential, the essential ingredients can
be identified by reviewing the known properties of the gas, liquid and solid
phases of HF. The simplest associated form of HF is the gas phase dimer,
(HF)$_2$, whose structure and rovibrational spectrum were first characterized
by the microwave spectra of Dyke, Howard and
Klemperer. \cite{Dyke1972,Howard1984} The equilibrium configuration of the
isolated dimer F-H$'\cdots$F-H$''$ is planar but bent, as a consequence of a
competition between dipolar and quadrupolar electrostatic interactions
\cite{Howard1984,Kolebrander1988,Barton1982} (whenever ambiguity is possible,
H$'$ denotes the hydrogen atom involved in the hydrogen bond, while H$''$ is
the other one). The atoms F-H$'\cdots $F form a nearly linear arrangement,
with H$'$ placed slightly off the FF axis, and with a distance between the
hydrogen-bonded fluorines $r_{{\rm FF}}\approx $ 2.7 \AA . The second hydrogen
atom forms an angle $\FFH \approx 115^{\circ}$ and the H--F bonds are slightly
stretched with respect to the
monomer. \cite{Howard1984,Huber1979,Kofranek1988}

The (HF)$_2$ bent dimer appears to be the basic structural motif of
all the best known associated forms of HF, namely the gas phase
(HF)$_6$ cyclic hexamer, \cite{Janzen1969} the low temperature
crystal \cite{Johnson1975} and even the
liquid. \cite{Deraman1985} Pairs of adjacent HF units in the
hexamer exhibit a bent arrangement similar to that of the dimer. The
same structure is found in crystalline deuterium fluoride (DF) which
is made up of infinite zig-zag chains of DF
units. \cite{Johnson1975} Interchain F--F distances ($\approx
3.2$ \AA ) are much larger than intrachain F--F distances ($\approx
2.5$ \AA ), an indication that interchain interactions are weaker than
the interactions between adjacent HF units within the chains. The very
small entropy difference between the solid and the liquid
\cite{Vanderzee1970} suggests that the liquid is largely
associated. Best fit analysis of dielectric constant data
\cite{Cole1973} and of Raman and IR spectra
\cite{Desbat1983} support this conclusion and indicate that
acyclic zig-zag chains are abundant in the liquid. Finally, the
neutron scattering measurements \cite{Deraman1985} of the
radial correlation function of liquid DF yield H--F and F--F average
neighbor distances consistent with a most probable local structure
similar to that of the other forms of HF.

Since all associated forms of HF share the (HF)$_2$ dimer as a common
structural unit, any satisfactory potential model should reproduce the
main features of the (HF)$_2$ potential energy surface. {\Ai} quantum
mechanical calculations have contributed significantly to clarify the
interactions present in the isolated dimer (HF)$_2,$ by determining
accurately its equilibrium properties
\cite{Peterson1995,Collins1995} and by mapping out the entire
potential energy
surface. \cite{Kofranek1988,Bunker1988,Bunker1990,Jensen1990}
Much of this work has been directed to obtain analytical models for
the potential surface, usually by fitting the numerical results to a
chosen functional
form. \cite{Kolebrander1988,Peterson1995,Bunker1988,Bunker1990,Jensen1990,Quack1990}
The analytical expressions of these models, which have been developed
mainly to study the rovibrational states of (HF)$_2 $,
\cite{Kolebrander1988,Barton1982,Jensen1990,Quack1990,Zhang1995}
are usually very complex and therefore not suitable for simple MD or
MC simulations.

Fortunately, as already mentioned, there is evidence that the main
features of the real molecular interactions may be approximately
reproduced by much simpler empirical models with effective
potentials. In fact, it has been long known
\cite{Howard1984,Kolebrander1988,Barton1982} that the bent
structure of the (HF)$_2$ dimer is dominated by the classical two-body
electrostatic interactions between the permanent multipole moments of
the monomer (mainly the dipolar and quadrupolar terms). The
incorporation of many-body polarization effects gives rather small
refinements on the predicted equilibrium angles,
\cite{Kolebrander1988} a finding consistent with the
experimental observation that the dipole moment of the dimer is only
weakly enhanced relative to that of the isolated
monomer. \cite{Dyke1972}

The effects of the environment on the HF molecules do not seem to be
large.  However, a comparison of the structures in the sequence
``monomer $\rightarrow $ dimer $\rightarrow $ hexamer $\rightarrow $
liquid $\rightarrow $ solid'' (as shown by Tables \ref{t:gas} and
\ref{t:condensed} below) evidences a tendency in which a larger degree
of association implies a small increase of the H--F bond distance,
together with a much larger decrease of the F--F distance (from 2.8
{\AA} in the gas phase dimer to 2.49 {\AA} in the
solid). Unfortunately, as stressed by R\"{o}thlisberger and
Parrinello, \cite{Rothlisberger1997} the available three-site
models
\cite{Klein1979,Cournoyer1984,Jedlovszky1997a,Jedlovszky1997b}
refer to {\it rigid} molecules and, consequently, they cannot
reproduce the relaxation of the interatomic distances in going from
the gas phase dimer to the condensed phases.

The previous observations suggest a relatively clear picture of the
ingredients required for a satisfactory potential model for HF. First,
a charge distribution is required which approximately reproduces the
first few multipolar moments of the HF molecule. Second, the H--F bond
cannot be considered as rigid and {\it intra}-molecular interactions
must be included to allow for the observed variations of the H--F and
F--F distances in the various aggregation forms of HF. Finally,
further atom--atom interactions must be introduced to model the
remaining non-Coulombic intermolecular forces. Having all this in
mind, we have tried to construct a new three-site model by
simultaneously using data on the gas, liquid and solid phases. For
this purpose: (a) the molecules are not rigid, but the H--F bond
length can vary, and, (b) the parameters are fitted to theoretical and
experimental data, including the {\ai} structure of the HF dimer,
\cite{Kofranek1988} the room temperature density of liquid DF
\cite{Deraman1985} and the experimental structure of solid DF
at 4 K. \cite{Johnson1975}

For the structure of the dimer we have used the {\ai} (HF)$_2$
potential energy surface developed by Bunker, Jensen, Karpfen,
Kofranek and Lishka
(BJKKL). \cite{Kofranek1988,Bunker1988,Bunker1990,Jensen1990}
The BJKKL calculations, in which the potential energy of the (HF)$_2$
complex has been determined for over 1000 different configurations,
represent the most complete and accurate scan of the energy surface to
date. These quantum mechanical results are in excellent agreement with
the experiments on gas phase (HF)$_2$.

The decision to add to the fit some data on solid and liquid HF (more
precisely, DF) has been taken because of the partial failure of our
preliminary models fitted only to the {\ai} surface. These models
reproduced well the zig-zag (HF)$_\infty $ chains characteristic of
the crystal, but gave totally wrong inter-chain distances and,
furthermore, did not agree with the experimental density of the
liquid. \cite{Deraman1985} This behavior was to be
expected. In fact, the (HF)$_2$ potential surface
\cite{Kofranek1988,Bunker1988,Bunker1990} only accounts for
the basic HF--HF interactions {\it within} a chain, and obviously
exclude the weak long-range interactions responsible for the distances
between different (HF)$_\infty $ chains. Since the density of the
liquid is affected by interactions between distant pairs of HF
molecules in relative orientations which are not sampled in the solid,
it is also understandable that only by fine-tuning the long range
potential it has been possible to reproduce the experimental density
of the liquid. Finally, it must be mentioned that the addition of data
on solid and liquid HF to the fit gave only a small deterioration of
the agreement with the {\ai} data on the dimer. This fact confirms
that solid and liquid data add information on regions of the potential
surface that are not sampled by the dimer.

\section{Methods and Calculations}

\subsection{Potential model} \label{ss:pot}

The potential model is represented by intra- and inter-molecular
parts:
\begin{equation}
V_{\text{HF}}^{\rm intra}(r_{\text{HF}})=D_e\left\{ 1-\exp[-\alpha
(r_{\text{HF}}-r_e)]\right\}^2,
\end{equation}

\begin{equation}
V_{AB}^{\rm inter}=V_{AB}^{\rm Coul}+V_{AB}^{\rm non-Coul},
\end{equation}

\begin{equation}
V_{AB}^{\rm Coul}=\sum_{i\in A} \sum_{j\in B} \frac{q_iq_j}{r_{ij}},
\end{equation}

\begin{equation}
V_{AB}^{\rm non-Coul}=A_{{\rm FF}}\exp(-B_{{\rm FF}}~r_{{\rm
FF}})-C_{{\rm FF}}~r_{{\rm FF}}^{-6}.
\end{equation}

BJKKL \cite{Bunker1990} fitted the intra-molecular part of
their own {\ai} surface with a Morse potential, eq. (1). Here $r_e$
represents the equilibrium H-F distance, $D_e$ the dissociation energy
and $\alpha $ is an effective range parameter. Because of their
simplicity and accuracy, the BJKKL functional form and parameter
values are adopted in this paper.

The Coulombic interactions between molecules $A$ and $B$ are modeled
through three point charges for each HF monomer, two at the nuclear
positions $\R_{{\rm H}}$ and $\R_{{\rm F}}$, and the third at a
position $\R_{{\rm X}}$ along the H--F bond. In eq. (3) $q_i$ and
$q_j$ are the fractional charges on the $i$th site of molecule $A$ and
$j$th site of molecule $B,$ respectively; $r_{ij} $ is the distance
between these sites. The motion of the site X is constrained so that
it remains at the same relative position along the bond,
\begin{equation}
\R_{{\rm X}}=\beta \R_{{\rm F}}+(1-\beta )\R_{{\rm H}},
\end{equation}

\noindent where $\beta $ is an adjustable parameter between 0 and
1. The charges are $+q$ at both the H and F nuclei, and $-2q$ at the
third site to preserve neutrality. This three-site charge model is
related to that of
Refs. \onlinecite{Klein1979,Cournoyer1984,Jedlovszky1997a,Jedlovszky1997b,Sun1992}.
By allowing for changes in the H--F bond length, the present model
effectively accounts for a part of the polarization effects. For solid
and liquid DF the Ewald's method
\cite{Born1954,Signorini1991,Allen1987} has been used to
ensure complete convergence of the Coulombic interactions (which have
an infinite range).

The remaining non-Coulombic part of the inter-molecular potential is
represented in a simplified way, using only a Buckingham ``$\exp\!-6$''
atom-atom interaction between the fluorines, eq. (4). This term is meant to
represent the interactions between the electronic clouds around two far away
atoms. Since the hydrogens in HF are essentially bare nuclei, it makes good
physical sense to avoid atom-atom interactions involving them. As a matter of
fact, no improvement in the quality of the fit is found by adding similar H--H
and H--F non-Coulombic interactions.

A rather well defined hierarchy of interactions may be identified in
the chosen potential model. The length of the HF monomer is solely
determined by the intra-molecular potential. The structure of the
dimer is also influenced by the position of the charge site X and by
the F--F equilibrium distance, i.e. by the position of the minimum of
the $\exp\!-6$ interaction between the fluorines. Finally, the charge
and the remaining properties of the $\exp\!-6$ model, mainly the
strength of the long range attractive term $C_{{\rm FF}}~r_{{\rm
FF}}^{-6}$, affect the structure and density of solid and liquid
HF. The presence of this hierarchy implies that small changes in the
charge and in the long range attraction can be compensated by the
remaining parameters to maintain the correct monomer and dimer
structures.

\subsection{Potential optimization}

The potential model contains five adjustable parameters, $\beta $,
$q$, $A_{{\rm FF}}$, $B_{{\rm FF}}$ and $C_{{\rm FF}}$, and three
parameters fixed at the {\ai} values, \cite{Bunker1990} $r_e$,
$\alpha $ and $D_e$.

In a first series of attempts, we fitted the present model (as well as
other preliminary ones) only to the {\ai} potential energy of the
dimer. The $\chi ^2$ deviation between the model and the {\ai} surface
was computed, for each given combination of parameter values, with the
same relative weights used by BJKKL. \cite{Bunker1990} The
$\chi^2$ was minimized by searching the parameter space with
Nelder-Mead simplex method. \cite{Press1986} The resulting
potential was then tested by computing some properties of the other
associated forms of HF. In particular, the liquid phase was studied by
isothermal-isobaric MD simulations, as described in the next
Subsection. As discussed in the introduction, these preliminary models
fitted only to the {\ai} surface gave unsatisfactory results, so that
it was decided to add more data to the fit.

For this purpose, the equilibrium geometries at $T=0$ K of the HF
dimer, hexamer and crystal were determined as a function of the
potential parameters by minimizing, with the WMIN program,
\cite{Busing1984} the total potential energy with respect to
the structural parameters. The deviations of the calculated geometries
from the {\ai} dimer structure \cite{Kofranek1988} and from
the experimental DF crystal structure \cite{Johnson1975} at 4 K
are then added to $\chi ^2$, with weights
subjectively chosen to make the contributions from the surface, the
dimer and the crystal roughly equal.

Since the new set of parameters, although more satisfactory, still did
not give the correct density of the liquid, it become necessary to
tune the long range interactions. In a further set of minimization
runs, a range of $q$ and $C_{{\rm FF}}$ values was searched, again
with Nelder-Mead method. The three remaining parameters $\beta $,
$A_{{\rm FF}}$ and $B_{{\rm FF}}$ where determined as a function of
$q$ and $C_{{\rm FF}}$ by fitting the {\ai} and crystal data. Each
complete set of five parameters was then used in a short MD simulation
to determine the equilibrium density of the liquid at 293 K and to add
to the $\chi^2$ the deviation from the relevant experimental
value. \cite{Deraman1985} This method, which involves two
nested fit procedures, was found to be reasonably efficient and
converged to a minimum in about fifty cycles.

The main problem encountered in the fit was the noise in the computed liquid
density due to insufficient MD equilibration with our computer time
constraints. To reduce this noise, the MD equilibration was done in parallel
with the potential optimization, by accepting after successful $\chi^2$
minimization cycles the final configuration of the MD run as the initial
configuration for the next run. With this strategy the current MD
configuration was always the one with the best potential parameters so
far. Since the parameters change rather slowly, the simulated system tended
to remain close to equilibrium. No structural data on the liquid, beside the
density, has been included in the fit. This rather drastic choice avoids the
repeated calculation of equilibrated radial correlation functions, which
would have required even longer MD runs then those needed for equilibrated
densities.

As a technical detail, it must be noticed that no attempt was done to
embed potential surface calculation, MD simulation, dimer and crystal
energy minimizations and the two nested Nelder-Mead procedures into a
single monolithic program, which would have been unmanageably
complex. Separate programs, calling each other as distinct processes
at the operating system level, \cite{Bourne1982} were used
instead. The two Nelder-Mead procedures, in particular, were actually
a single program invoking a second copy of itself. No special
changes were required for the surface, MD and energy minimization
programs.

The optimal set of parameter values is shown in Tab. \ref{t:pot} and
represents a compromise among the best results that can obtained
separately for the dimer, the crystal and the liquid. As usual, no
special physical significance should be attributed to the potential
parameters. In fact, because of the possible compensation among
different terms in the potential model, alternative slightly different
sets of parameters might have been used. The model may be simply
regarded as a tool to reproduce the observed data and predict new
results.

\subsection{Molecular Dynamics} \label{ss:md}

The MD calculations employed 500 {\it deuterium} fluoride (DF)
molecules, in a cube with periodic boundary conditions, and, using
Andersen's isothermal-isobaric (NPT) method
\cite{Andersen1980,Brown1984,Fox1984} simulated a liquid
sample in contact with a heat bath and subject to a hydrostatic
pressure of 1 atm ($\approx 10^{-4}$ GPa). The simulated liquid was
obtained by melting and equilibrating at 293 K an initial crystalline
configuration.  The behavior of the system as a function of
temperature was then determined by raising or lowering the bath
temperature in steps of $10$ K. Each temperature was maintained for at
least 5 ps for equilibration and further 5 ps for analysis. The
equations of motion were integrated using the velocity Verlet
algorithm, \cite{Allen1987,Brown1984,Fox1984,Verlet1967} with
a time step of 0.25 fs. As previously described, each DF molecule
carried three point charges, at D, F and X sites. Ciccotti, Ferrario
and Ryckaert method for linear constraints \cite{Ciccotti1982}
has been used to maintain each massless X charge at a fixed fraction
$\beta $ of the DF bond.

\section{Results}

The most important properties calculated with the present potential
model for the monomer, dimer, planar (HF)$_n$ rings, hexamer, crystal
and liquid forms of HF (or DF) are compared in Tables \ref{t:gas},
\ref{t:rings} and \ref{t:condensed} with the available experimental
and {\ai} data. As described in section \ref{ss:md}, the properties of
the liquid have been determined through MD calculations, while the
equilibrium geometries at 0 K of the other forms of HF are found by
minimizing the potential energy.

\subsection{Monomer, dimer and cyclic polymers of HF}

The excellent results (Tab. \ref{t:gas}) for the equilibrium bond
length, dissociation energy and spectroscopic parameters of the
monomer, which all depend only on the parameters fixed to the BJKKL
values, \cite{Bunker1990} indicate that the Morse model
accurately reproduces the main features of the true intra-molecular
potential. The spectroscopic parameters $\nu_e$ (harmonic frequency)
and $x_e$ (anharmonicity constant) for the energy levels of a Morse
oscillator, $E_n=h\nu_e[(n+\frac{1}{2})-x_e(n+\frac{1}{2})^2]$, are
directly obtained from $D_e$ and $\alpha $ through $\nu_e=\alpha
\sqrt{D_e/\mu}$ and $x_e=h\nu_e/4D_e$, where $\mu $ is the reduced HF
mass. \cite{Child1984}

The multipole moments computed from our three-charges model (taking
the molecular center of mass as origin) are very close to the
experimental and {\ai} moments of the
monomer. \cite{Bunker1988,Gray1984} The dipole and quadrupole
values follow the trend of the {\ai} results and therefore are
slightly underestimated with respect to the experimental data. The
octupole and hexadecupole moments are also in reasonable agreement
with the {\ai} calculations. This overall agreement is an indication
of a good match between the electrostatic interactions in real HF and
in the model.

The computed minimum energy structure of the dimer (Tab. \ref{t:gas})
compares well with the experimental and {\ai}
data. \cite{Howard1984,Kofranek1988,Bunker1988,Pine1986} The
experimental F-F distance is excellently reproduced, as well as both
the angles ${\HFF}$ and ${\FFH}$ of the bent equilibrium
configuration. Moreover, the increased length of the HF molecule in
going from the monomer to the dimer, and the slight length difference
between the two HF intramolecular bonds, \cite{Kofranek1988}
are well predicted. Since these length changes were the primary reason
for allowing non-rigid molecules, such a behavior must be considered
very satisfactory. Unfortunately, in spite of the excellent dimer
geometry, the dimerization energy is clearly underestimated. This
drawback was not completely unexpected. In fact, in real HF and in
quantum mechanical models dimerization is accompanied by hydrogen
bonding, which is not explicitly incorporated in the present classical
treatment.

As a further test of the potential, we have computed the equilibrium
geometry of the planar (HF)$_n$ rings, with $n=2,3\cdots8$. As shown
in Tab.  \ref{t:rings}, the structural parameter computed for planar
rings, with $C_{nh}$ symmetry, are in good agreement with the
available {\ai} results. \cite{Kofranek1988,Karpfen1990}
Though the model systematically underestimates the {\ai} binding
energies, it nevertheless reproduces correctly the relative stability
of the different structures. The smallest ring, which is the cyclic
dimer, has a binding energy of $\Delta E=-3.18$ kcal/mole, and is thus
substantially less stable than the bent dimer (Tab. \ref{t:gas}).

For cyclic (HF)$_3$ the binding energy for each hydrogen bond,
$-\Delta E/3$, is slightly less that the binding energy of the bent
dimer. The bond stabilization energy, $-\Delta E/n$, increases for
larger rings, up to the hexamer, and then decreases again. The
particularly favorable stability of the (HF)$_n$ rings with
$n\approx6$ is readily understood by noticing that for these rings the
${\HFF}$ and ${\FFH}$ angles (Tab. \ref{t:rings}) are close to those
of the bent dimer (Tab. \ref{t:gas}). Planar (HF)$_n$ rings must
satisfy the geometric constraint ${\FFH}-{\HFF}=\alpha_n$, where
$\alpha_n=180^\circ-360^\circ/n$ is the inner angle of the $n$-sides
regular polygon. The hexamer, for which $\alpha_n=120^\circ$, can be
obtained by joining essentially undeformed bent dimers, and is the
most stable structure.

The hexamer can be stabilized even further by allowing for non planar
structures. We find that the stablest structure has a non-symmetric
``chair'' shape, with average bond lengths and angles
(Tab. \ref{t:gas}) which compare well with the available experiments
\cite{Janzen1969} and which are almost identical to those
found in the dimer. A ``boat'' structure slightly above in energy, at
$-4.96$ kcal/mole, is also found, with lengths and angles close to
those of the ``chair'' structure.

\subsection{Crystal and liquid DF}

Low temperature crystals of DF are orthorhombic, space group $Cmc2_1$
($C_v^{12}$), with four molecules per unit cell on the $\sigma_v$
plane sites. \cite{Johnson1975} The minimum energy structure
computed for the DF crystal (see Tab. \ref{t:condensed}) is close to
the experimental structure at low
temperature. \cite{Johnson1975} The discrepancies in the
lengths of the cell axes partially compensate each other, yielding a
density only slightly smaller than the experimental one. The increased
H-F intra-molecular lengths with respect to both monomer and dimer,
and the F-F distance smaller than in the dimer, are well reproduced.

In Fig. \ref{f:density} the densities predicted by the present model
over the whole temperature range of liquid HF (which, at atmospheric
pressure, goes from the freezing point at $-83$ $^{\circ}$C up to the
boiling point at $19.75$ $^{\circ}$C) are compared with the
experimental data. The experimental densities show a nearly linear
dependence on $T.$ The data are from two sources, covering different
temperature ranges. \cite{Simons1932,Sheft1973} The slight
vertical shift between the two data sets is due to the experimental
uncertainties. It is very satisfactory to note that, although our
potential model has been fitted only to a single density at $293$ K,
the straight line corresponding to its predictions is somewhat
vertically shifted, but has essentially the same slope as the
experimental density. This slope is not reproduced by the HFC, HF-JV1
and HF-JV2 models, \cite{Cournoyer1984,Jedlovszky1997a,Jedlovszky1997b} 
since their density straight lines (obtained by
joining the corresponding results, which, unfortunately, are available
only at $203$ and $273$ K) intersect the experimental curve.

Another pleasant feature of the present model, also shown in
Fig. \ref{f:density}, is that at $303$ K and above the average density
continued to decrease over the whole MD analysis period. At each
$T<303$ K, the density oscillated around an equilibrium value. In our
opinion, this behavior indicates that the simulated liquid boils at
some point in the range $293\div 303$ K, in good agreement with the
experimental normal boiling point of HF ($T_b=292.9$ K).

With regard to the internal energy $U,$ our MD results exhibit a
nearly linear dependence on $T.$ However, their absolute values
(Tab. \ref{t:condensed}) are underestimated with respect to the
available experimental data \cite{Jedlovszky1997a} (a similar
drawback is also present in the HF3 model \cite{Klein1979}).

Fig. \ref{f:radial} reports the partial pair radial correlation
functions $g_{ij}$(r) computed for the liquid at 293 K, together with
the real space function $d(r)\equiv 4\pi \rho_m r\left[ G(r)-1\right]
$. Here $\rho_m$ is the molecular number density, and $G(r)$ is a
composite (or total) pair correlation function, obtained by adding the
three partial pair correlation functions with the appropriate nuclear
weights \cite{Deraman1985} and then by convoluting the sum
with the same experimental resolution function used in the neutron
scattering experiments. \cite{Deraman1985} The resulting
theoretical prediction for $d(r)$ is compared in Fig. \ref{f:radial},
with the corresponding neutron diffraction data for liquid DF at 293
K, which are the only available real space
data. \cite{Deraman1985}

The first peak at $r\approx 0.95$ {\AA} is due to the intra-molecular
H--F bond distance and here the agreement between simulation and
experiment is excellent. The second and third peak of the experimental
$d(r)$ occur at $r\approx 1.6$ {\AA} and $r\approx 2.55$ {\AA}, and
correspond to the hydrogen-bond $r_{{\rm HF}}$ inter-molecular
distance and to the $r_{{\rm FF}}$ separation,
respectively. Unfortunately, the model (with the present choice of
parameters) fails to reproduce these peaks and the complex liquid
structure at longer distances. The reason of such a shortcoming may be
found by comparing the partial pair correlation functions $g_{ij}(r)$
of the present model Fig. (\ref{f:radial}) with the analogous {\ai} MD
results of R\"{o}thlisberger and
Parrinello. \cite{Rothlisberger1997} The position of the first
inter-molecular peak of all our $g_{ij}(r)$ is shifted toward larger
$r $ values (a similar trend occurs in the polarizable HF-JV2 model
\cite{Jedlovszky1997b}). In addition, the height of the
hydrogen-bond peak of $g_{{\rm HF}}(r)$ is nearly half the correct
value, and the height of the first peak of $g_{{\rm HH}}(r)$ is also
underestimated.

It is to be recalled that the HF3 potential \cite{Klein1979}
reproduces the three principal peaks of the experimental data
\cite{Deraman1985} and gives the best performance for $d(r)$
among the available models. This quite good agreement was obtained by
modeling the hydrogen-bond interaction with a Morse
term. \cite{Klein1979} To our knowledge, no other empirical
model takes explicitly into account the hydrogen bond.

Before concluding this Section, it may be useful to summarize the
performances of the best available models for liquid HF. The HF3
potential gives, as already mentioned, quite good agreement with
neutron diffraction structural data at the normal boiling point (293
K), \cite{Deraman1985} but systematically fails to reproduce
thermodynamics: \cite{Klein1979} the predicted internal energies are
largely underestimated (i.e., their absolute values are too small) and
the pressures in MD calculations at constant volume are several
kilobars too high, indicating that the model system is less strongly
bound than real HF. In comparison with HF3, the HFC model
\cite{Cournoyer1984} yields a slightly better thermodynamics but a
slightly worse structure for the liquid, with charges corresponding to
a dipole moment enhanced with respect to the monomer. Then, the
predictions of HF-JV1 for the liquid phase \cite{Jedlovszky1997a} are
not very different from the HFC ones, but HF-JV1 fails completely to
reproduce the properties of the isolated dimer. Finally, whereas
HF-JV1 works reasonably well only at room temperature, the polarizable
model potential HF-JV2 represents a true improvement
\cite{Jedlovszky1997b} as regards the predicted density at low
temperature, i.e. at $203$ K. However, problems with the pair
correlation functions are encountered also for HF-JV2.

\section{Conclusions}

We have shown that a simple potential model suitable for MD or MC
simulations can reproduce, quantitatively or semiquantitatively, many
physical properties of the hydrogen fluoride over the whole set of its
solid, liquid and gaseous associated forms. To cover such a wide range
of environmental conditions, it has been necessary to consider a
molecular model with variable bond length. The present investigation
confirms the plausibility of assuming the dimer as the basic
structural unit, but also stresses the need of fitting the potential
to a set of experimental and theoretical data which includes
information on the condensed phases.

In the liquid, the correct trend of the density is very encouraging, but the
underestimated energies and the problems with the radial distribution
functions indicate that some physical effect is still misrepresented by the
model. Unfortunately, it is not likely that adding data on the liquid
structure could improve the results with the current type of model (without
an hydrogen bond term). In fact, our experience with the fit shows that those
few parameter sets which give better liquid structures were incompatible with the
gas and crystal data. We believe that the essential shortcoming of the model
is the neglect of any {\it explicit} representation for the hydrogen bond,
which cannot be reduced to purely electrostatic multipole interactions. Such
an explicit modeling of the hydrogen bond is also lacking in most other
three-site potentials for HF, with the exception of the HF3
model. \cite{Klein1979} Although the hydrogen bonding has a quantum
mechanical origin, an approximate classical treatment is however possible and
may have significant consequences, as seen from the rather good structural
results for the HF3 potential. \cite{Klein1979} The inclusion of potential
terms representing the hydrogen bond interactions appears therefore the next
necessary step for more accurate HF models. More data on the liquid,
including at least the radial correlation function, need to be incorporated
in the fit.

In conclusion, our model cannot be considered as a definitive solution
of the problem, but can be seen as a significant step towards a really
satisfactory potential for MD or MC simulations. It has the merit of
pointing out the importance of a variable molecular length and of the
hydrogen bond. Moreover, it shows that the strategy of a simultaneous
fit to data covering all the associated forms of HF can be successful
and should be considered as the appropriate way to fully accomplish
the difficult task of finding a potential model for such a strongly
associating system.

\acknowledgments Work done with funds from MURST (Ministero
dell'Universit\`a e della Ricerca Scientifica e Tecnologica) through
the INFM (Istituto Nazionale di Fisica della Materia), from CNR and
from the University of Bologna (``Finanziamento speciale alle
strutture''). We thank Bunker, Jensen, Karpfen, Kofranek, and Lishka
for providing their {\ai} data.

\bigskip
\begin{figure}
\caption{Density of liquid DF as a function of temperature $T$. The
center and half width of the error bars represent the average and
standard deviation of the MD results, respectively, sampled over the
last 5 ps at each bath temperature. The arrow at 303 K spans the range
of densities sampled at this $T$, and indicates that the density was
decreasing through the whole analysis period. The filled circle
represents the experimental DF density, \protect\cite{Deraman1985}
whereas the empty circles represent HF measurements,
\protect\cite{Simons1932,Sheft1973} multiplied by the DF/HF mass ratio
to obtain consistent units.}
\label{f:density}
\end{figure}

\begin{figure}
\caption{Upper panel: total neutron pair radial correlation function
$d(r)$ for liquid DF (continuous line, simulation; dots, experiment
\protect\cite{Deraman1985}). Lower panel: $g_{ij}(r)$ partial pair
correlation functions from the simulation. The curves for $g_{\rm
HH}(r)$ and $g_{\rm FF}(r)$ have been vertically displaced.}
\label{f:radial}
\end{figure}

\bigskip
\begin{table}[ht]
\caption{Potential parameters. The meaning of the parameters is described in
section \protect\ref{ss:pot}. The Morse parameters $r_e$, $\alpha$ and $D_e$
have been fixed to the values of the corresponding parameters ($c_1$, $c_3$
and $k_8^{000}/8\pi^{3/2}$, respectively) of the BJKKL fit to the {\ai}
surface. \protect\cite{Bunker1990}}
\begin{tabular}{ll}
Monomer equilibrium length, $r_e$ ($a_0$)            & 1.73727 \\
Monomer Morse parameter, $\alpha$ ($a_0^{-1}$)       & 1.174   \\
Monomer dissociation energy, $D_e$ (Hartree)         & 0.22306 \\
Position of the X site, $\beta = {\rm XF/HF}$        & 0.16245 \\
Charge, $q$, $-2q$, $q$ on F, X, H, $q$ ($e$ units)  & 0.59456 \\
Buckingham parameter, $A_{\rm FF}$ (kcal/mol)        & 167017  \\
Buckingham parameter, $B_{\rm FF}$ (\AA$^{-1}$)      & 4.148   \\
Buckingham parameter, $C_{\rm FF}$ (\AA$^6$kcal/mol) & 547.9   \\
\end{tabular}
\label{t:pot}
\end{table}

\bigskip
\begin{table}[ht]
\caption{Properties of gas-phase HF monomer and polymers. The {\ai} hexamer
data \protect\cite{Karpfen1990} are for a planar ring, whereas the model results are for the
minimum energy ``chair'' structure.}
\begin{tabular}{llllll}
Monomer                                        & Model   & Experimental & Ref. & {\Ai} & Ref. \\
\hline
HF equilibrium distance, $r_e$ (\AA)           &  0.9193 & 0.91680 & [\onlinecite{Huber1979}] & 0.9194 & [\onlinecite{Kofranek1988}] \\
dissociation energy, $D_e$ (kcal/mole)         &   140.0 & 141.6   & [\onlinecite{Huber1979}] & 141.2  & [\onlinecite{Feller1997}] \\
harmonic frequency, $\nu_e$ (cm$^{-1}$)        & 4120.16 & 4138.32 & [\onlinecite{Huber1979}] & 4135~~ & [\onlinecite{Kofranek1988}] \\
anharmonicity constant, $x_e\nu_e$ (cm$^{-1}$) &~~~86.69 &~~~89.88 & [\onlinecite{Huber1979}] &~~~90.1 & [\onlinecite{Feller1997}] \\
fundamental frequency,						      			
$\Delta\nu_{0\rightarrow 1}$ (cm$^{-1}$)       & 3946.78 & 3961.42 & [\onlinecite{Blanc1994}] &        & \\
dipole moment,      $\mu_{z}$    (D)           &   1.772 & 1.826   & [\onlinecite{Gray1984}]  & 1.7728 & [\onlinecite{Bunker1988}] \\
quadrupole moment,  $Q_{zz}$ (D \AA)           &   2.122 & 2.36~   & [\onlinecite{Gray1984}]  & 2.3048 & [\onlinecite{Bunker1988}] \\
octupole moment,    $\Omega_{zzz}$ (D \AA$^2$) &   1.894 &         &             & 1.7327 & [\onlinecite{Bunker1988}] \\
hexadecupole moment, $\Phi_{zzzz}$ (D \AA$^3$) &   1.658 &         &             & 1.87~~ & [\onlinecite{Gray1984}] \\
\\
Dimer                                & Model   & Experimental  & Ref.       & {\Ai}   & Ref. \\
\hline
HF$'$ distance (\AA)                 & 0.9270  &               &              & 0.9236  & [\onlinecite{Kofranek1988}] \\
HF$''$ distance (\AA)                & 0.9222  &               &              & 0.9220  & [\onlinecite{Kofranek1988}] \\
FF distance (\AA)                    & 2.6850  & 2.72$\pm$0.03 & [\onlinecite{Howard1984}] & 2.7919  & [\onlinecite{Kofranek1988}] \\
${\HFF}$ angle (degrees)             &~~~7.52  & ~~10$\pm$6    & [\onlinecite{Howard1984}] &~~~6.81  & [\onlinecite{Kofranek1988}] \\
${\FFH}$ angle (degrees)             & 110.86  & ~117$\pm$6    & [\onlinecite{Howard1984}] & 114.45  & [\onlinecite{Kofranek1988}] \\
dimerization energy $\Delta E$ (kcal/mole) & $-$4.03 & $-$4.56 & [\onlinecite{Pine1986}]   & $-4.32$ & [\onlinecite{Bunker1988}] \\
\\
Hexamer                              & Model   & Experimental    & Ref.       & {\Ai}  & Ref. \\
\hline
HF distance (\AA)                    & 0.9333  & 0.973$\pm$0.009 & [\onlinecite{Janzen1969}] & 0.948  & [\onlinecite{Karpfen1990}] \\
FF distance (\AA)                    & 2.6217  & 2.535$\pm$0.003 & [\onlinecite{Janzen1969}] & 2.475  & [\onlinecite{Karpfen1990}] \\
${\HFF}$ angle (degrees)             &~~~6.15  &                 &             & ~~~2.4  & [\onlinecite{Karpfen1990}] \\
${\FFH}$ angle (degrees)             & 111.24  & 104             & [\onlinecite{Janzen1969}] & 117.6  & [\onlinecite{Karpfen1990}] \\
binding energy $\Delta E$ (kcal/mole) & $-4.98$ & $-7.20$     & [\onlinecite{Redington1981}] & $-8.3$ & [\onlinecite{Karpfen1990}] \\
\end{tabular}
\label{t:gas}
\end{table}
\noindent

\bigskip
\begin{table}[ht]
\caption{Interatomic distances, angles, and bond stabilization energies of
(HF)$_n$ planar rings with $C_{nh}$ symmetry.}
\begin{tabular}{ccccccccccl}
    & \multicolumn{4}{c}{Model} &~~~& \multicolumn{4}{c}{{\Ai}} \\
$n$ & HF    & FF    & $\FFH$    & $\Delta E/n$ && HF    & FF    & $\FFH$    & $\Delta E/n$ & Ref.\\
    & (\AA) & (\AA) & (degrees) & (kcal/mol)   && (\AA) & (\AA) & (degrees) & (kcal/mol)   & \\
\hline
2   & 0.9247 & 2.6950 & ~49.34 & $-1.59$ && 0.9223 & 2.796 & ~54.23 & $-3.30$ & [\onlinecite{Kofranek1988}] \\
3   & 0.9321 & 2.6568 & ~81.37 & $-3.79$ && 0.932~ & 2.616 & ~83.6  & $-5.1$  & [\onlinecite{Karpfen1990}] \\
4   & 0.9335 & 2.6246 & 101.04 & $-4.71$ && 0.943~ & 2.522 & 101.6  & $-7.2$  & [\onlinecite{Karpfen1990}] \\
5   & 0.9326 & 2.6338 & 114.41 & $-4.90$ &&  \\					
6   & 0.9315 & 2.6474 & 123.90 & $-4.92$ && 0.948~ & 2.475 & 122.4  & $-8.3$  & [\onlinecite{Karpfen1990}] \\
7   & 0.9308 & 2.6599 & 131.02 & $-4.89$ &&  \\
8   & 0.9302 & 2.6688 & 136.50 & $-4.86$ &&  \\
\end{tabular}
\label{t:rings}
\end{table}

\bigskip
\begin{table}[ht]
\caption{Properties of solid and liquid DF}
\begin{tabular}{llll}
Solid DF at 4.2K               & Model           & Experimental    & Ref. \\
\hline
HF distance (\AA)              & 0.933           & 0.95$\pm$0.02   & [\onlinecite{Johnson1975}] \\
FF distance (\AA)              & 2.662           & 2.50$\pm$0.01   & [\onlinecite{Johnson1975}] \\
${\FFH}$ angle (degrees)       & 113.1           & 116.6$\pm$1.0   & [\onlinecite{Johnson1975}] \\
unit cell axis $a$ (\AA)       & 3.29            & 3.31            & [\onlinecite{Johnson1975}] \\
unit cell axis $b$ (\AA)       & 4.50            & 4.26            & [\onlinecite{Johnson1975}] \\
unit cell axis $c$ (\AA)       & 5.31            & 5.22            & [\onlinecite{Johnson1975}] \\
density (g/cm$^3$)             & 1.77            & 1.89            & [\onlinecite{Johnson1975}] \\
binding energy $\Delta E$ (kcal/mole) & $-7.41$  & $-10.7$         & [\onlinecite{Zunger1975}] \\
\\								
Liquid DF at 293K              & Model           & Experimental    & Ref. \\
\hline								
HF distance (\AA)              & 0.930$\pm$0.022 & 0.958$\pm$0.002 & [\onlinecite{Deraman1985}] \\
FF distance (\AA)              & 2.777$\pm$0.169 & 2.56            & [\onlinecite{Deraman1985}] \\
binding energy $\Delta E$ (kcal/mole) & $-3.70$  & $-6.93$         & [\onlinecite{Cournoyer1984}] \\
density (g/cm$^3$)             & $1.02\pm0.03$   & 1.0106          & [\onlinecite{Deraman1985}] \\
boiling temperature $T_b$ (K)  & $\le303$        & 292.90          & [\onlinecite{Sheft1973}] \\
\end{tabular}
\label{t:condensed}
\end{table}

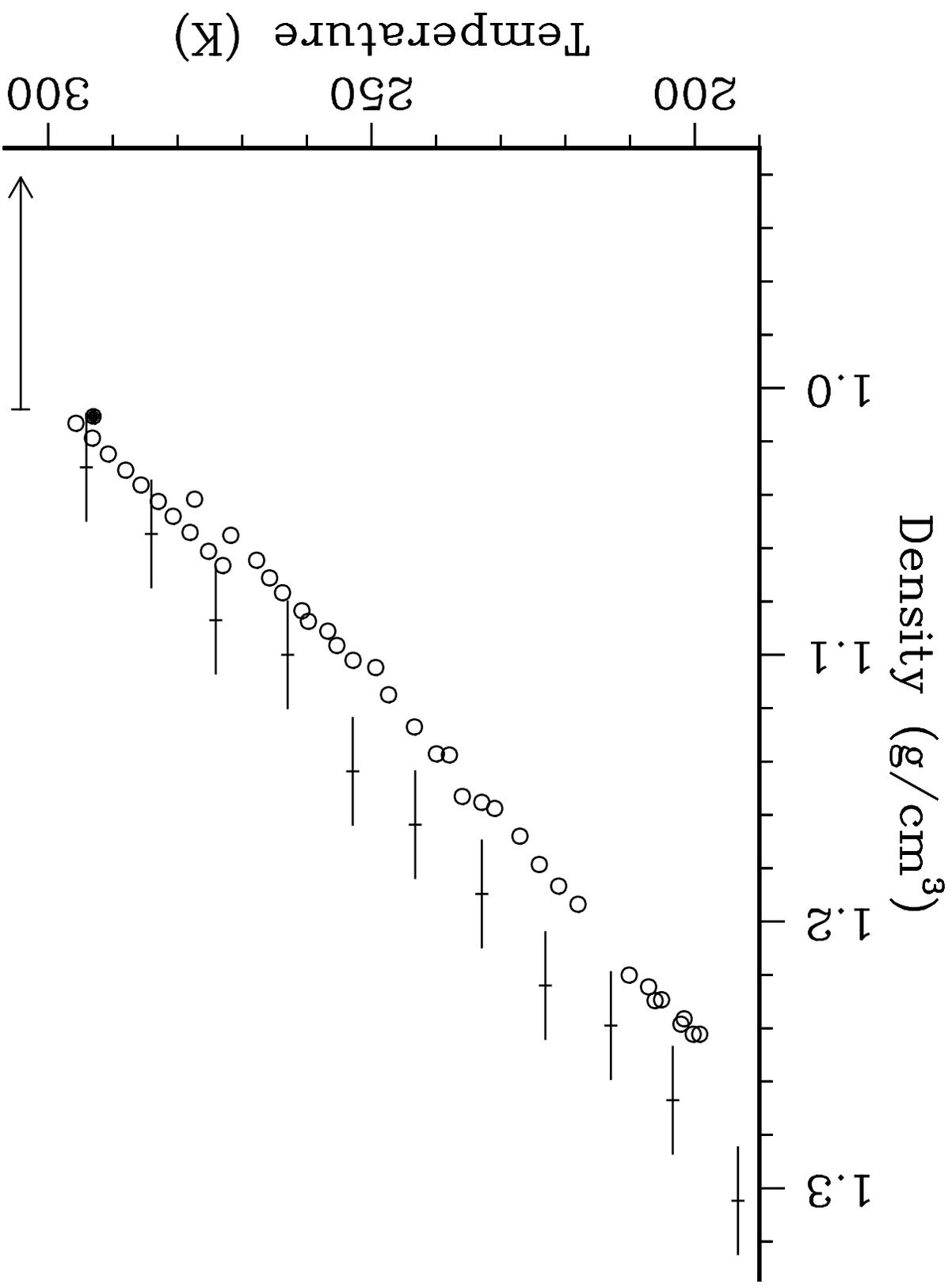
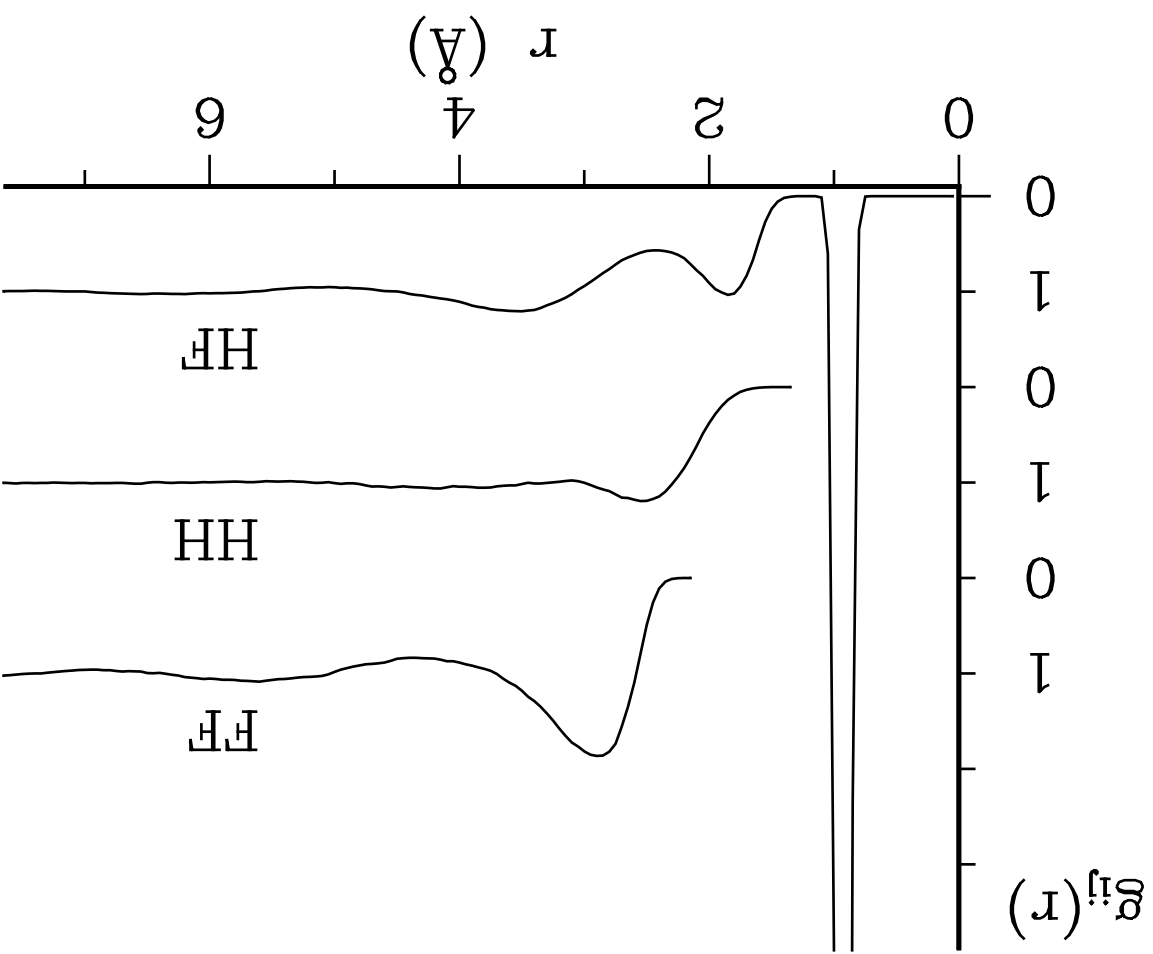
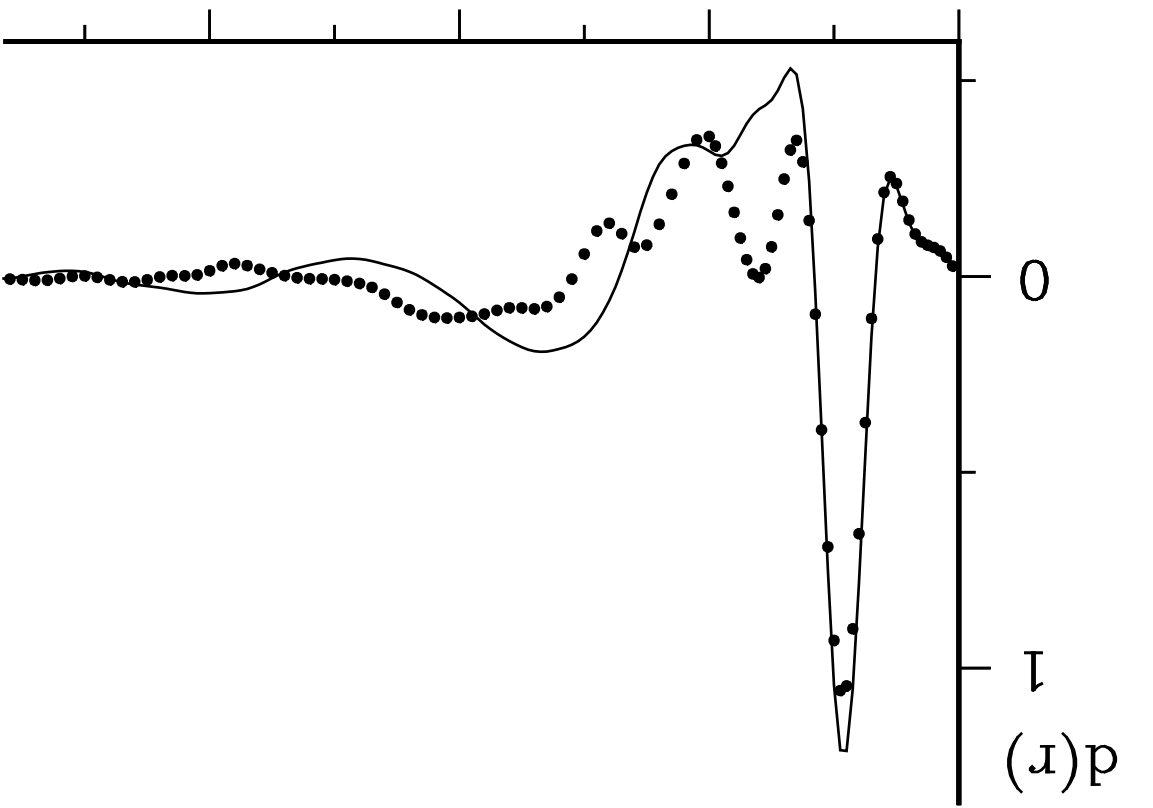
\end{document}